DRAFT WHITE PAPER

# A call for regularly updated review/survey articles: "Perpetual Reviews"

David L. Mobley[1] and Daniel M. Zuckerman[2]

[1]Departments of Pharmaceutical Sciences and Chemistry, University of California, Irvine, dmobley@uci.edu

[2]Department of Computational and Systems Biology, University of Pittsburgh, ddmmzz@pitt.edu

**Abstract**: We advocate the publication of review/survey articles that will be updated regularly, both in traditional journals and novel venues. We call these "perpetual reviews." This idea naturally builds on the dissemination and archival capabilities present in the modern internet, and indeed perpetual reviews exist already in some forms. Perpetual review articles allow authors to maintain over time the relevance of non-research scholarship that requires a significant investment of effort. Further, such reviews published in a purely electronic format without space constraints can also permit more pedagogical scholarship and clearer treatment of technical issues that remain obscure in a brief treatment.

**We advocate perpetual reviews -- regularly updated review papers which provide a current snapshot of research within their scope:** We believe the scholarly category of "review papers" needs revisiting in view of the latest technology by adding a new category of "perpetual reviews" which are kept updated. Current review papers, while extremely valuable, quickly are superseded or made obsolete by the latest work in a given area. While review papers can in some cases maintain considerable impact over the years, later researchers must by necessity consult more recent literature to ensure they are not missing something important which falls within the scope of the original review. In our view, this is far from ideal. The authors of a review paper ought to be able to update it to include new references within the scope of the paper, and potentially expand it to address new topics within the same overall scope. This constitutes the core idea of perpetual reviews.

Several projects exist that support perpetual reviews. One is LivingReviews.org, which directly implements the approach advocated here. Note that "living reviews" is a trademark owned by the Max Planck Society, and hence we avoid the phrase in this revision of our initial submission. Another project is The Net Advance of Physics, which includes links to (mostly static) reviews and tutorials. Scholarpedia.org takes a related but slightly different approach, and is discussed below.

**Perpetual reviews are updateable, versioned, and updated regularly:**

Perpetual reviews should meet at least two key criteria. First, they will allow perpetual updates by their authors as desired by the author (subject perhaps to reasonable constraints on frequency and length by the journal). Second, they will be versioned (and all versions archived in an accessible way), so that citations can refer to the specific version cited and readers can easily access both the exact version cited as well as the current version. Additionally, there also ought to be reasonable expectations about update frequency. For example, a review paper which is not updated for a period longer than a year[1] ought to be considered a "standard review" (and noted as such) unless its authors again begin updating it regularly.

Essentially, perpetual reviews ought to be regarded as snapshots of the current state of knowledge within a given area, updated regularly by their author who is an expert in that area.

In our view, perpetual reviews will not be updated forever - there will be a natural time of retirement. This may occur in some cases because the author simply ceases updating them, for personal or professional reasons. But in other cases, the author might decide to declare an article "retired" by publishing new work(s) in the area which would replace the original work. In this case, the perpetual review ought to be updated to reflect these new references, and to clearly indicate that it is no longer a perpetual review but reflects a snapshot of the area at the date of its last update.

**Perpetual reviews suggest a category of review papers without space constraints:**

The idea of perpetual reviews leads naturally to a category of more pedagogical review papers which operate without space constraints. Specifically, such articles could have a longer, more explanatory format, and feature pedagogical overviews, extensive explanations without page limits, and individual discussions of referenced articles (available by hyperlink and/or in appendices or supporting documents). These could be better suited than traditional reviews to providing students or new researchers an introduction to a particular area of research. While these might become lengthy and unwieldy if not organized properly, the authors could deal with this by (i) by retiring a particular perpetual review and dividing it naturally into new perpetual reviews in relevant sub-areas, as needed, or (ii) organizing the material so that finer-grained explanations appear as appendices or via hyperlinks.

**A "virtual journal" of perpetual reviews may be in order:**

In order to encourage perpetual reviews, we are considering beginning a virtual journal of such reviews, which would link to and highlight perpetual reviews within our field, wherever they are published. We hope this might encourage adoption of the idea of perpetual reviews by a range of journals and authors. In our view, virtual reviews themselves could be published in a variety of forums, including in the online archives of traditional journals, or in alternative publishing formats such as arXiv.org or escholarship.org, or perhaps even via GitHub. The key requirements for a perpetual review would remain the same - permanently updateable, versioned, and updated regularly.  See the related effort: [The Net Advance of Physics](#).

---

[1] In many fields, scholarship moves enough over the space of a year that any review paper covering that area ought to require updates. However, other fields may differ and each ought to set its own expectations in this regard.

With such a virtual journal, we would hope to encourage healthy debate. For example, we would be delighted to be able to link to reviews offering multiple perspectives on the same subject, even if these views differ.

We note that a virtual journal with versioned archiving could sidestep some of the technical hurdles associated with the permanency of the [DOI](#) (digital object identifier) system used in traditional journal publishing.  The DOI system, in a way, maps digital publishing to the older paper system - for example, requiring errata rather updates.

We welcome feedback and expressions of interest regarding this idea from the community.

**Our motivation is in part the apparent inefficiency of the current system of reviews:**

Consider an example review on some topic and published in a fairly prominent journal. In our experience, when we publish such a review, inevitably we almost immediately become aware of new literature we should have cited - either as a result of the publication of new work within the scope of our review, or because readers contact us and make us aware of other old work which ought to have been cited since it was within our scope. Typically, within perhaps even a month after publication of a review paper, we could add a substantial number of references which would considerably improve the work. Currently, review papers provide no mechanism for such updates. Certainly, errata can be used to correct mistakes, but most possible updates to reviews do not involve correction of mistakes.

It is natural that authors should want to be able to update their work rather than simply discarding it as time goes on. Currently, this sometimes seems to be dealt with in our specific fields by authors publishing new review papers in different journals after several years have passed; these often cover much of the same material, but re-written and updated (and sometimes leaving out old material which they feel was adequately covered in a previous review). What this means is that if we enter a new field and want to find out the current state of understanding in the field, we cannot get by with just reading the latest review papers from the leaders in the field. Typically we must also read earlier review papers, since these also might include additional important material, as well as any new papers published in the field after the review papers were written. Additionally, pedagogical reviews are few and far between.

In our view, the current system would be dramatically improved by allowing authors to publish perpetual reviews which are targeted to specific audiences. Some review papers could be deliberately pedagogical, without length limits, and the authors could expand them as much as needed to provide an appropriate introduction to students or researchers entering an area. Some of these might naturally migrate to other formats, such as becoming textbook chapters or similar. Other review papers would be targeted more towards advanced researchers in a particular area, those wanting to know an author's current perspectives in that area or those wanting to see the current snapshot of the latest work in that area. Perpetual reviews, whether technical or pedagogical, would provide researchers with an always-up-to-date view of the material they are interested in, from the perspective they want. To us, this seems vastly superior to our current system.

**While websites and encyclopedias can be valuable, we believe current scholarly incentives often work against these:**

Academic incentives, as we understand them, do not properly reward the creation of non-journal material, such as websites.

One of the authors (DLM) has experience with Alchemistry.org, a community website intended to cover free energies. The idea was to get the community to contribute to a Wikipedia-like site which would provide a number of things, including:

- An introduction to free energy calculations
- A summary of current (always updated) best practices in the area, with references and discussion
- A background of the theory and methods
- Tutorials
- Worked examples/samples

In some sense, the vision for the site encompasses some of the same areas that would be covered by suitable perpetual reviews of the area.

While the site itself is fairly extensive and contains a good deal of valuable material, it has been hard to engage the community in contributing. We believe that in part, this is because at least in academia, our productivity is often measured in terms of publications. And contributing to a website is not currently recognized as a publication. So academic incentives work against contributing to a community website; the time is often seen as "better spent" on writing material for more traditional publication. Some of these disincentives have been noted elsewhere, as reasons why academics may not spend much effort disseminating knowledge via Wikipedia and similar sources [http://www.scholarpedia.org/article/Help:Frequently_Asked_Questions].

We believe perpetual reviews will help to deal with this, since they should still be citeable as traditional publications. Authors and their evaluators may still have challenges, since in many cases a work is not considered in a case for academic promotion unless it was done during the period being reviewed. However, if a review is a perpetual review, that could be grounds for an exception. And even if not, authors will be able to point to how many new citations a perpetual review received during that period.

Scholarpedia.org has an interesting model: it intends to be a peer-reviewed, open-access encyclopedia, where articles are written by experts in a given area and updates are then curated by an expert in the area (often the author or one of the authors). Given that these articles can be updated perpetually, and they are versioned, they can in some sense be considered perpetual reviews. However, they do differ somewhat from what we envision. First, these are formatted more as encyclopedia articles than reviews, meaning that often they may not target quite the same audience as a review paper. Second, updates can come from contributors other than the author (and are moderated by a curator, who may or may not be the author). So authorship may become less clear - and harder for evaluators to consider - as an article's lifespan extends.

As noted above, the [Living Reviews](#) journals already implement perpetual reviews.

**Longer term: Could there be a role for living/perpetual research articles?**

While somewhat tangential to the present discussion, it is interesting to contemplate the idea of "living research articles." Imagine the convenience of being able to study a group's multi-year efforts on a single topic by reading a single cohesive monograph! Needless to say, this idea would need to be aligned with academic incentives, or *vice versa*. The notion of authorship would need to be revisited, certainly, although such a review is perhaps overdue already. We leave aside this idea for another day.

**In summary, we believe perpetual reviews provide a path forward within our current system of scholarly credit**

We discussed the idea of perpetual reviews: easily updateable, regularly updated, and versioned review articles. Perpetual reviews, as conceived here and building on precedents, will be fully electronic and hence permit greater explanatory discussion without narrow page-limit constraints. We believe such reviews are needed by the community and would provide a significant and immediate improvement over what is currently possible. Hopefully, traditional academic journals will begin offering authors an opportunity to publish perpetual reviews online. Otherwise, other publication means may need to be sought.

We suggested a virtual journal of perpetual reviews, which will link to perpetual reviews wherever they are published (subject to oversight of an editorial board) including both traditional and non-traditional means of publication. This may help the idea of perpetual reviews to gain traction.

**Acknowledgements**
We appreciate constructive discussions with Ivet Bahar, Ken Dill, Bill Jorgensen, Terry Stouch, and David Zuckerman, as well as informative correspondence from Frank Schulz and Norman Hugh Redington.